\documentclass[twoside]{dis07}
\usepackage[latin1]{inputenc}
\usepackage[dvips]{graphicx,epsfig,color}
\usepackage{wrapfig,rotating}
\usepackage{amssymb,amsmath,array}

\begin{document}
\title{Parton Distributions for LO Calculations}

\author{R.S. Thorne$^{1}$\thanks{Royal Society University Research Fellow},
A. Sherstnev$^2$ and C. Gwenlan$^1$
%
% DO NOT MODIFY THE FOLLOWING '\vspace' ARGUMENT
\vspace{.3cm}\\
%
% Addresses and institutions (remove "1- " in case of a single institution)
1- Department of Physics and Astronomy, University College London, WC1E 6BT, UK
% Remove the next three lines in case of a single institution
\vspace{.1cm}\\
2- Cavendish Laboratory, University of Cambridge, Cambridge, CB3 0HE, UK\\
}

%***********************************************************************
% END OF AUTHORS INFORMATION AREA
%***********************************************************************

\maketitle

\begin{abstract}
  We present a study of the results obtained combining LO partonic 
matrix elements with different orders of partons distributions. These are 
compared to the {\em best} prediction using NLO for both 
matrix elements and parton distributions. 
The aim is to determine which parton distributions are most appropriate 
to use  in those cases where only LO matrix elements are available, e.g. as in 
many Monte Carlo generators. Both LO and NLO parton distributions have faults 
so a modified {\em optimal} LO set is suggested.  
\end{abstract}

The combination of the order of the parton distribution function 
(pdf) and the 
accompanying matrix element is an important issue \cite{url}. 
It has long been known that 
LO pdfs in some regions of $x$ and $Q^2$ are qualitatively different to 
NLO (and NNLO) pdfs (see \cite{DIS06}) due to important missing 
corrections in splitting functions or coefficient functions for 
structure   
\begin{wrapfigure}{r}{0.48\columnwidth}
\vspace{-0.65cm}
\centerline{\includegraphics[width=0.40\columnwidth]{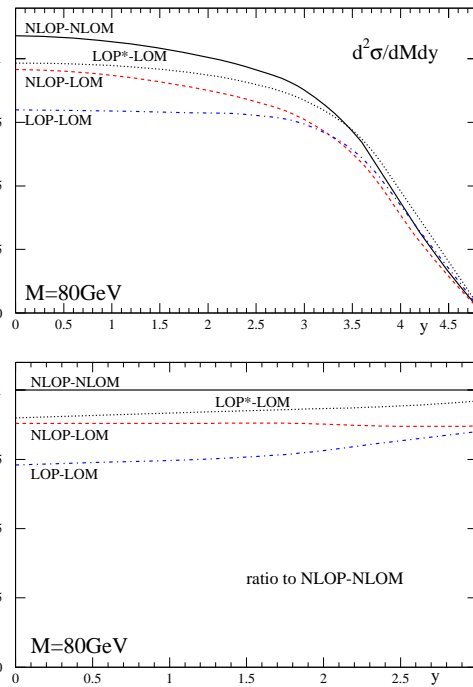}}
\vspace{-0.5cm}
\caption{Drell-Yan distribution at the LHC.}\label{Fig:lhc}
\vspace{-0.4cm}
\end{wrapfigure}
functions which are used to determine the pdfs.
Nevertheless, LO pdfs are
usually thought to be the best choice for use with LO matrix elements, such 
as those available in many LO Monte Carlo programs, though all such results 
should be treated with care.
Recently it has been suggested that NLO 
pdfs may be more appropriate \cite{Joey},
since NLO cross-section corrections are often small. There has  
already been an investigation of the use of NLO 
pdfs for the underlying event \cite{Field}. 
There is a big difference in the results when using CTEQ6L and CTEQ6.1M
pdfs \cite{CTEQ6} due to the changes in the gluon, though
agreement can be reached by significant retuning. This will affect 
predictions for other quantities.

In this article we address the differences in predictions obtained for a 
variety of physical quantities combining different  pdfs with 
LO matrix elements. In each case NLO pdfs combined 
with NLO matrix elements represent the best prediction -- the {\em truth}. 
We interpret the 
features of the results and investigate how a best set of pdfs for use 
with LO matrix elements may be obtained.  

First, let us recall how LO pdfs tend to differ from NLO pdfs. 
The most marked differences are for light quarks at high $x$ 
and the gluon distribution at low $x$. The coefficient functions for 
structure functions have $\ln(1-x)$ enhancements at higher perturbative order.
This means the high-$x$ quarks are smaller as the order increases. 
The quark-gluon splitting 
function  $P_{qg}$ develops a small-$x$ divergence at NLO (with further 
$\ln(1/x)$ enhancements at higher orders), so the small $x$ gluon needs to be 
bigger at LO in order to fit structure function evolution. Indeed, for 
$Q^2\sim 1-2$ GeV$^2$ the NLO gluon is flat or valence-like at small $x$
while at LO it grows much more quickly. These features show up 
in cross-sections.  

Let us start with the simple example of Drell-Yan production of 
vector-bosons from quark-antiquark annihilation as a function of rapidity at 
fixed invariant mass = $80$GeV, i.e. appropriate for $W,Z$ production, at 
the LHC.  The NLO correction \cite{DYNLO} for Drell-Yan 
production is quite significant, 
being  positive for all rapidity and roughly $12\%$ in this case.  
The absolute predictions and the ratios of LO matrix element results to 
the {\em truth} are show in Fig. \ref{Fig:lhc}. We see that indeed  
we are nearer to the {\em truth} with LO matrix elements and NLO pdfs 
\cite{MRST04}
than  LO matrix elements and LO pdfs \cite{MRSTLO}. 
Both LO ME results are too small, but NLO pdfs are 
closer and a better shape. 
LO pdfs and the LO matrix elements have the wrong shape being low 
at central rapidity but increasing at high rapidity where the high-$x$ quark 
is enhanced at LO. Hence, NLO pdfs seem more appropriate here.

\begin{wrapfigure}{r}{0.48\columnwidth}
\vspace{-0.7cm}
  \centerline{\includegraphics[width=0.40\columnwidth]{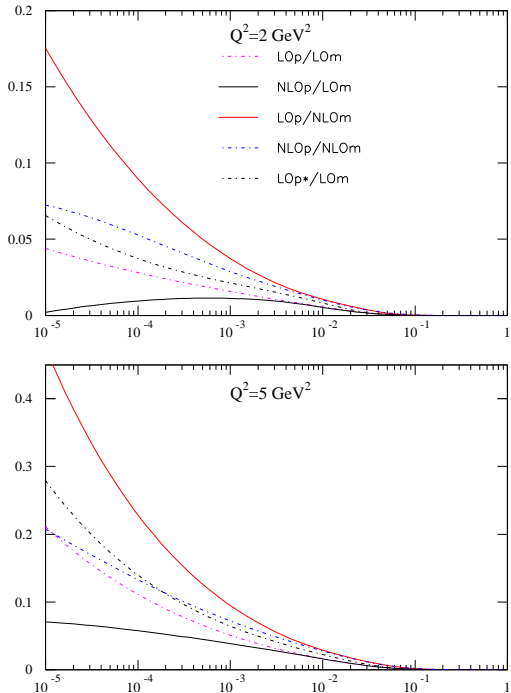}}
\vspace{-0.5cm}
  \caption{The structure function $F_2^c(x,Q^2)$.}
\label{Fig:charm}
\vspace{-0.5cm}
\end{wrapfigure}

However, there is a small $x$ counter-example. 
Consider the production of charm in low $Q^2$ DIS at HERA energies. The NLO
matrix elements {\cite{nlocalc} 
contain a divergence at small $x$ which is not present at LO.
Using NLO pdfs the LO matrix element result is well
below the {\em truth} at low scales and the shape is totally wrong,
as seen in Fig. \ref {Fig:charm}. 
The LO gluon is very large at small $x$ since it has been extracted 
with missing $P_{qg}$ enhancement at small $x$, and this compensates the 
missing small $x$ divergence in the matrix element --  
LO pdfs and LO matrix elements are more sensible with a  
compensation between failings in both. Using LO pdfs and NLO matrix 
elements the result is extremely large since there is a double counting of the 
small $x$ divergence. 

From these two examples we can conclude that 
sometimes NLO pdfs are better to use if only the LO matrix 
elements are known, and we
can get significant problems with the size and shape if LO pdfs are used. 
However, we can be completely wrong, particularly at small $x$, if we use
NLO pdfs due to {\em zero}-counting of small-$x$ terms.  
Can we find some optimal definition of pdfs which have most 
desirable features? In order to make progress we 
need to better understand the difference between LO and NLO pdfs. 
The missing higher order
terms in $\ln(1-x)$, $1/x$ and $\ln(1/x)$ in coefficient functions
and/or evolution leads to pdfs at LO which are bigger at $x \to 1$ 
and at $x \to 0$ in order to compensate. However,  
from the momentum sum rule there are then not enough partons to go around,
and enhancements in some regions lead to depletion in other regions, 
particularly 
quarks at $0.001\leq x \leq 0.1$. This  leads to a bad global fit at LO 
\cite{MRSTLO} -- 
partially compensated by  the LO extraction of $\alpha_S(M_Z^2)$ being 
$\sim 0.130$ to help speed evolution, and explains the underestimate of the 
Drell-Yan production at the LHC at more central rapidities.

This obvious source of problems has lead to a suggestion \cite{Sjostrand} 
that relaxing the momentum sum rule could make LO pdfs rather more 
like NLO pdfs in the regions where they are normally too small. The 
resulting pdfs would still be bigger than NLO where 
necessary, i.e. the high-$x$ quarks and low-$x$ gluon, but would not be 
depleted elsewhere. It is also useful to use the NLO definition 
of $\alpha_S$. Because of quicker running at NLO, the LO and NLO couplings 
with the same value of $\alpha_S(M_Z^2)$ are very different at lower scales
where DIS data used in global fits exists. 
Near $Q^2=1$ GeV$^2$ the NLO coupling with 
$\alpha_S(M_Z^2)=0.120$ is similar to the LO coupling with 
$\alpha_S(M_Z^2)=0.130$. Hence, 
the use of the NLO coupling helps alleviate the discrepancy between 
pdfs at different orders. Indeed, the NLO coupling already used in 
some CTEQ LO pdfs \cite{CTEQ6} and 
in LO Monte Carlo generators. Relaxing momentum conservation for input 
pdfs  and using the NLO definition of  
 $\alpha_S$ does dramatically improve the quality of the LO global fit,
$\chi^2=3066/2235$ for the standard LO fit becoming 
$\chi^2=2691/2235$,  with a big improvement in the comparison to HERA data.
The momentum carried by the input pdfs goes up to $113\%$.  
Using the NLO definition the value of $\alpha_S(M_Z^2)=0.121$.

\begin{wraptable}{r}{0.34\columnwidth}
\vspace{-0.3cm}
  \centerline{\begin{tabular}{|c|c|c|}
      \hline
      parton & matrix  &     $\sigma$ (mb) \\
          &  element   &     \\
      \hline
      NLO  & NLO  &  41.5     \\
      LO  & LO & 24.8     \\
      LO*  & LO  & 34.8   \\
      NLO  & LO  &  16.8  \\
      \hline
  \end{tabular}}
\vspace{-0.3cm}
  \caption{$\sigma(b\bar b)$ totals.}
  \label{Tab:table}
\vspace{-0.4cm}
\end{wraptable}

The modified pdfs, which we denote by LO*, become much more 
similar to NLO pdfs, in particular at small $x$ the LO* quark 
distributions evolve as quickly as at NLO and are similar for 
for $x \sim 0.001-0.01$. Similarly $g(x,Q^2)$ is significantly 
bigger at LO* than at LO, and much bigger than NLO at small $x$. This will 
help when used with LO matrix elements for 
gluon-gluon initiated processes (e.g. 
Higgs production) where $K$-factors are often much greater than 
unity. We now look at the LO* pdfs in our first two examples,
see Figs. \ref{Fig:lhc} and \ref{Fig:charm}. 
For Drell-Yan production at the LHC the LO* pdfs lead to shape of 
comparable quality as the NLO pdfs and the normalization is better.
For the charm structure function comparing all possibilities 
the LO* pdfs and LO matrix elements are indeed 
closest to the {\em truth}.

\begin{wrapfigure}{r}{0.6\columnwidth}
\vspace{-0.4cm}
  \centerline{\includegraphics[width=0.60\columnwidth]
{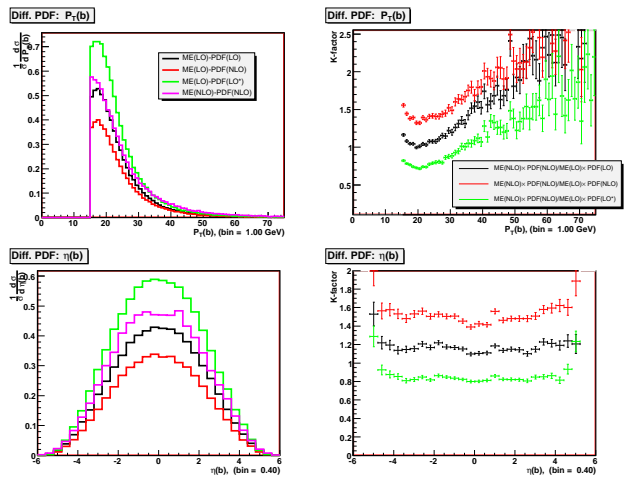}}
\vspace{-0.4cm}
  \caption{The $b$ cross section at the LHC.}\label{Fig:b}
\vspace{-0.2cm}
\end{wrapfigure}

There is a similar conclusion for hadro-production of $b$ quarks 
at LHC which probes low $x \sim 0.001$. We consider $\sigma(b\bar b)$ where 
the initial $b$ quark has  
$p_T>10$ GeV and $\vert \eta \vert \leq 5$ using a NLO event generator
\cite{MCatNLO} and LO calculations \cite{LObottom}. 
The total cross-sections are 
shown in Table. 1. The NLO pdfs and LO matrix element are clearly worst. 
We also 
illustrate final state $p_T$ and rapidity distributions in Fig. \ref{Fig:b} 
with $p_T>18$ GeV
for the $b$ quark after showering. Again the best absolute predictions with 
LO matrix elements uses LO* pdfs and the worst NLO pdfs. 
However, in this case there is always a problem with the shape as function 
of $p_T$. The NLO matrix element has a large positive effect at high $p_T$ 
and very high $\eta$. It is impossible for any parton shape to account 
for all NLO corrections.

We also look at very high-$E_T$ jet production \cite{jets} 
at the LHC in Fig. \ref{Fig:jets}. Ignoring the lowest $E_T$, where 
hadronization and underlying event  and possibly small $x$ physics are an 
issue, the LO and LO* pdfs with the LO
matrix elements are of comparable quality to NLO pdfs.
\begin{wrapfigure}{r}{0.5\columnwidth}
\vspace{-0.3cm}
\centerline{\includegraphics[width=0.5\columnwidth]{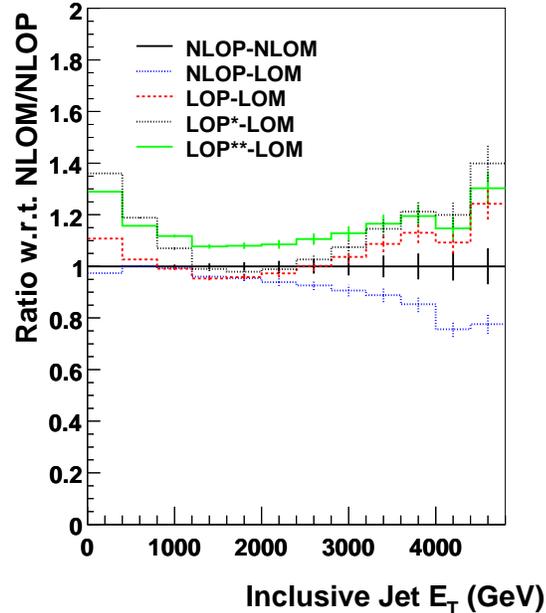}}
\vspace{-0.6cm}
\caption{The dijet cross-section at the LHC.}\label{Fig:jets}
\vspace{-0.3cm}
\end{wrapfigure} 
LO and NLO pdfs when 
combined with LO matrix elements produce results which 
deviate in opposite directions at high $E_T$. 
Also shown is the prediction from LO** pdfs, which have 
an additional constraint for a harder high-$x$ gluon (this being small at LO 
even with momentum conservation relaxed). This changes the results 
little compared to the use of the LO* pdfs. 

Hence, we can conclude that a fixed prescription of either LO or NLO pdfs 
with LO matrix elements or LO Monte Carlo generators will lead each to 
incorrect results in some cases. To try to improve this situation we have 
suggested an optimal set of pdfs for LO calculations, the LO* pdfs, 
which are essentially LO but with various modifications to make their 
features more NLO-like. These seem to work reasonably well and happen to 
achieve some 
of the features obtained by modifying pdfs in a process dependent fashion
for use in Monte Carlo 
generators/resummations discussed in e.g. \cite{Mrenna}. More study is 
underway. However, sometimes 
NLO matrix element corrections qualitatively change the features of 
the predictions for a physical process, regardless of how careful one is with 
pdfs, since new types of partonic process open up. This must
always be borne in mind, and accounted for if possible.

\begin{footnotesize} 

\end{footnotesize}

\end{document}